\documentclass[a4paper,nofootinbib,showpacs,twocolumn]{revtex4}
\usepackage{graphicx}

\begin{document}

\title{Temperature Dependence of Paramagnetic Critical Magnetic Field
in Disordered Attractive Hubbard Model}

\author{E.Z. Kuchinskii$^1$, N.A. Kuleeva$^1$, M.V. Sadovskii$^1$$^,$$^2$}

\affiliation{$^1$Institute for Electrophysics, Russian Academy of Sciences, 
Ural Branch, Ekaterinburg 620016, Russia\\
$^2$M.N. Mikheev Institute for Metal Physics, Russian Academy of Sciences, 
Ural Branch, Ekaterinburg 620108, Russia}


\begin{abstract}

Within the generalized DMFT+$\Sigma$ approach we study disorder effects in
the temperature dependence of paramagnetic critical magnetic field $H_{cp}(T)$
for Hubbard model with attractive interaction. We consider the wide range of
attraction potentials $U$ -- from the weak coupling limit, when superconductivity
is described by BCS model, up to the limit of very strong coupling, when
superconducting transition is related to Bose -- Einstein condensation (BEC) of
compact Cooper pairs. The growth of the coupling strength leads to the rapid
growth of $H_{cp}(T)$ at all temperatures. However, at low temperatures
paramagnetic critical magnetic field $H_{cp}$ grows with $U$ much slower, than
the orbital critical field, and in BCS limit the main contribution to the
upper critical magnetic filed is of paramagnetic origin. The growth of the
coupling strength also leads to the disappearance of the low temperature region
of instability towards type I phase transition and Fulde -- Ferrell -- Larkin --
Ovchinnikov (FFLO) phase, characteristic for BCS weak coupling limit.
Disordering leads to the rapid drop of $H_{cp}(T)$ in  BCS weak coupling limit,
while in BCS -- BEC crossover region and BEC limit $H_{cp}(T\to 0)$ dependence
on disorder is rather weak. Within DMFT+$\Sigma$ approach disorder influence
on $H_{cp}(T)$ is of universal nature at any coupling strength and related only
to disorder widening of the conduction band. In particular, this leads to the
drop of the effective coupling strength with disorder, so that disordering
restores the region of type I transition in the intermediate coupling region.

\end{abstract}


\maketitle


\section{Introduction}

In the weak coupling region and for the weak disorder the upper critical magnetic
field of a superconductor is determined by orbital effects and usually is much
lower than the paramagnetic limit. However, the growth of the coupling strength
and disordering lead to the rapid growth of the orbital $H_{c2}$ possibly
overcoming the paramagnetic limit.

In this paper we study the behavior of paramagnetic critical field in the region
of very strong coupling of electrons of the Cooper pair and in the crossover
region from BCS -- like behavior for the weak coupling to Bose -- Einstein
condensation (BEC) in the strong coupling region \cite{NS}, taking disorder into
account (including the strong enough).

The simplest model to study the BCS -- BEC crossover is Hubbard model with
attractive interaction. Most successful approach to the studies of Hubbard model,
both to describe the strongly correlated systems in the case of repulsive
interactions and to study the BCS -- BEC crossover for the case of attraction,
is the dynamical mean -- field theory (DMFT) \cite{pruschke,georges96,Vollh10}.

In recent years we have developed the generalized DMFT+$\Sigma$ approach to
Hubbard model \cite{JTL05,PRB05,FNT06,PRB07,UFN12,HubDis,LVK16}, which is quite
effective for the studies of the influence of different external (outside those
taken into account by DMFT) interactions. This DMFT+$\Sigma$ method was used by
us in Refs. \cite{JETP14,JTL14,JETP15} to study the disorder influence on the
temperature of superconducting transition. In particular, for the case of semi --
elliptic initial density of states, adequate to describe three -- dimensional
systems, it was demonstrated that disorder influence on the critical temperature
(in the whole region of interaction strengths) is related only to the general
widening of the initial conduction band (density of states) by disorder
(the generalized Anderson theorem). In Ref. \cite{JETP17_Hc2}, using the
combination of the Nozieres -- Schmitt-Rink approximation and DMFT+$\Sigma$
in attractive Hubbard model we have analyzed the influence of disordering on the
temperature dependence of the orbital upper critical field $H_{c2}(T)$ both for
the wide region of coupling strengths $U$, including the BCS -- BEC crossover
region, and in the wide region of disorder up to the vicinity of Anderson
transition. Both the growth of the coupling strength and disorder lead to the
rapid growth of $H_{c2}$, leading in the BEC -- limit to unrealistically high
values of $H_{c2}(T\to 0)$, significantly overcoming the paramagnetic limit.

In this work we perform the detailed analysis of disorder influence on the
temperature dependence of paramagnetic critical magnetic field of a superconductor
for the wide range of coupling strengths $U$, including the BCS -- BEC crossover
region and the limit of very strong coupling.

It is well known, that in BCS weak coupling limit paramagnetic effects
(spin splitting) lead to the existence of a low temperature region at the phase
diagram of a superconductor in magnetic field, where paramagnetic critical field
$H_{cp}$ decreases with further lowering of the temperature. This behavior
signifies the instability leading to the region of type I phase transition,
where also the so called Fulde -- Ferrell -- Larkin -- Ovchinnikov (FFLO) phase
may appear \cite{S-Gam_Sarma,FFLO1,FFLO2} with Cooper pairs with finite momentum
$\bf q$ and spatially periodic superconducting order parameter.
In the following we limit ourselves to the analysis of type II transition and
homogeneous superconducting order parameter, allowing us to determine the
border of instability towards type I transition in BCS -- BEC crossover
and strong coupling regions at different disorder levels. The problem of
stability of FFLO phase under these conditions is not analyzed here.

\section{Hubbard model within DMFT+$\Sigma$ approach in Nozieres -- Schmitt-Rink
approximation}

We are considering the disordered Hubbard model with attractive interaction,
taking into account spin -- splitting by external magnetic field $H$, and
described by the Hamiltonian:
\begin{equation}
H=-t\sum_{\langle ij\rangle \sigma }a_{i\sigma }^{\dagger }a_{j\sigma
}+\sum_{i\sigma }\epsilon _{i}n_{i\sigma }-U\sum_{i}n_{i\uparrow
}n_{i\downarrow }-\mu_B H\sum_{i\sigma }\sigma n_{i\sigma },  
\label{And_Hubb}
\end{equation}
where $t>0$ -- is transfer amplitude between nearest neighbors, $U$ -- is the
onsite Hubbard attraction,
$n_{i\sigma }=a_{i\sigma }^{\dagger }a_{i\sigma }^{{\phantom{\dagger}}}$ -- is
electron number operator on a given site, $a_{i\sigma }$ ($a_{i\sigma }^{\dagger}$) --
electron annihilation (creation) operator, $\sigma = \pm 1$,
$\mu_B=\frac{e\hbar}{2mc}$ -- Bohr magneton, and local energies $\epsilon _{i}$
are assumed to be independent and random on different lattice sites.
We assume Gaussian distribution for energy levels $\epsilon _{i}$ at a given
site:
\begin{equation}
\mathcal{P}(\epsilon _{i})=\frac{1}{\sqrt{2\pi}\Delta}\exp\left
(-\frac{\epsilon_{i}^2}{2\Delta^2}
\right)
\label{Gauss}
\end{equation}
Distribution width $\Delta$ represents the measure of disorder, and Gaussian
random field of energy levels (independent on different lattice sites) produces
``impurity'' scattering, which is analyzed within the standard approach, based
on calculations of the averaged Green's functions.

The generalized DMFT+$\Sigma$ approach \cite{JTL05,PRB05,PRB07,FNT06,UFN12}
extends the standard dynamical mean field theory (DMFT
\cite{pruschke,georges96,Vollh10}  by addition of an ``external'' self -- energy
$\Sigma_{\bf p}(\varepsilon)$ (in general case momentum dependent) due to any
kind of interaction outside the DMFT, which gives an effective calculation
method both for single -- particle and two -- particle properties \cite{HubDis,PRB07}.
This approach conserves the standard system of self -- consistent DMFT
equations \cite{pruschke,georges96,Vollh10}, with ``external'' self -- energy
$\Sigma_{\bf p}(\varepsilon)$ being recalculated at each iteration step using
some approximate scheme, corresponding to the type of additional interaction,
while the local Green's function of DMFT is ``dressed'' by
$\Sigma_{\bf p}(\varepsilon)$ at each step of the standard DMFT procedure.

In the problem of disorder scattering under discussion here \cite{HubDis,LVK16}
for ``external'' self -- energy we are using the simplest (self -- consistent
Born) approximation, neglecting diagrams with ``crossing'' interaction lines due
to impurity scattering. Such an ``external'' self -- energy remains momentum
independent (local).

To solve the single -- impurity Anderson problem of DMFT in this paper,
as in our previous works, we use quite efficient method of numerical
renormalization group (NRG) \cite{NRGrev}.

In the following we assume the ``bare'' conduction band with semi -- elliptic
density of states (per unit cell with lattice parameter $a$ and single spin
projection), which gives a good approximation for three -- dimensional case:
\begin{equation}
N_0(\varepsilon)=\frac{2}{\pi D^2}\sqrt{D^2-\varepsilon^2},
\label{DOSd3}
\end{equation}
where $D$ defines the half -- width of the conduction band..

In Ref. \cite{JETP15} we have shown, that in DMFT+$\Sigma$ approach in the
model with semi -- elliptic density of states all the effects of disorder on
single -- particle properties reduce only to widening of conduction band
by disorder, i.e. to the replacement $D\to D_{eff}$, where $D_{eff}$ -- is the
effective ``bare'' band half -- width in the absence of correlations ($U=0$),
widened by disorder:
\begin{equation}
D_{eff}=D\sqrt{1+4\frac{\Delta^2}{D^2}}.
\label{Deff}
\end{equation}
The ``bare'' (in the absence of $U$) density of states, ``dressed'' by disorder
\begin{equation}
\tilde N_{0}(\xi)=\frac{2}{\pi D_{eff}^2}\sqrt{D_{eff}^2-\varepsilon^2},
\label{tildeDOS}
\end{equation}
remains semi -- elliptic also in the presence of disorder.

It is necessary to note, that in other models of the ``bare'' band disorder
not only widens the band, but also changes the form of the density of states.
In general, there is no complete universality of disorder influence on single --
particle properties, which reduces to the replacement $D\to D_{eff}$.
However, in the limit of strong enough disorder the ``bare'' band becomes
almost semi -- elliptic and this universality is restored \cite{JETP15}.

All calculations in the present paper, as in our previous works, were
performed for rather typical case of quarter -- filled band
(electron number per lattice site n=0.5).

To analyze superconductivity for the wide range of pairing interactions $U$,
following Ref. \cite{JETP15}, we use Nozieres -- Schmitt-Rink approximation
\cite{NS}, which allows qualitatively correct (though approximate) description
of BCS -- BEC crossover. In this approach, to determine the critical temperature
$T_c$ (in the absence of $H$) we use \cite{JETP15} the conventional BCS weak
coupling equation, but the chemical potential of the system $\mu$ for different
values of $U$ and $\Delta$ is determined from DMFT+$\Sigma$ calculations, i.e.
from the standard equation for the number of electrons in conduction band,
which allows us to find $T_c$  for the wide range of model parameters,
including the BCS -- BEC crossover region, as well as for different levels of
disorder. This reflects the physical meaning of Nozieres -- Schmitt-Rink
approximation: in the weak coupling region transition temperature is controlled
by the equation for Cooper instability, while in the strong coupling limit
it is determined as BEC temperature, which is controlled by chemical potential.
It was demonstrated, that such an approach guarantees the correct interpolation
between the limits of weak and strong couplings, including also the effects of
disorder \cite{NS,JETP14,JETP15}. In particular, in Refs.  \cite{JETP14,JETP15}
it was shown, that disorder influence on critical temperature $T_c$ and single --
particle characteristics (e.g. density of states) in the model with semi --
elliptic ``bare'' density of states is universal and is reduced only to the
changes of the effective bandwidth.

\section{Main results}

In the framework of Nozieres -- Schmitt-Rink approach the critical temperature
in the presence of spin -- splitting of electron level in external magnetic
field (and neglecting the orbital effects) or paramagnetic critical magnetic
field $H_{cp}$ at temperatures $T<T_c$ is determined by the following BCS --
like equation:
\begin{equation}
1=\frac{U}{4}\int_{-\infty}^{\infty}d\varepsilon 
\frac{\tilde N_0(\varepsilon)}{\varepsilon -\mu}
\left(
th\frac{\varepsilon -\mu -\mu_BH_{cp}}{2T}+
th\frac{\varepsilon -\mu +\mu_BH_{cp}}{2T}
\right),
\label{Hcp}
\end{equation}
where the chemical potential $\mu$ for different values of $U$ and $\Delta$ is
determined from DMFT+$\Sigma$ -- calculations, i.e. from the standard equation
for the number of electrons in conduction band. The general derivation of
Eq. (\ref{Hcp}) in the presence of disorder is given in the Appendix.
Note that Eq. (\ref{Hcp}) is derived from the exact Ward identity and remains
valid even in the case of strong disorder, including the vicinity of Anderson
transition. Eq. \ref{Hcp}) explicitly demonstrates, that all disorder effects on
$H_{cp}$ are reduced to the renormalization of the initial density of states by
disorder, so that for the case of initial band with semi -- elliptic density of
states disorder influence on $H_{cp}$ is universal and is only due to the band
widening by disorder, i.e. to the replacement $D\to D_{eff}$.

\begin{figure}
\includegraphics[clip=true,width=0.45\textwidth]{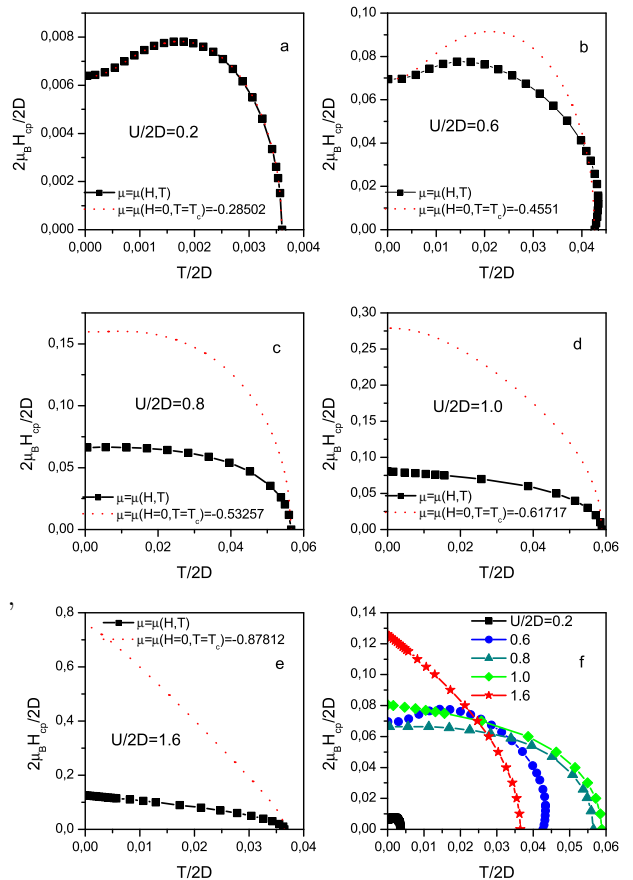}
\caption{Dependence of paramagnetic critical magnetic field on temperature for
different values of coupling strength. Dashed lines were obtained neglecting the
dependence of chemical potential on temperature and magnetic field at given $U$.}
\label{fig1}
\end{figure}

In Fig.\ref{fig1} we show the temperature dependence of paramagnetic critical
magnetic field for different values of coupling strength. Chemical potential
entering Eq. (\ref{Hcp}) is, in general, dependent not only on the coupling
strength, but also on the values of magnetic filed and temperature.
In Figs.\ref{fig1} (a-e), for the sake of comparison, dashed lines show the
results of calculations with chemical potential taken at $H=0$ and $T=T_{c}$
for the given value of $U/2D$, while continuous curves with symbols represent
the results of full calculations with $\mu=\mu(H,T)$.

In the weak coupling limit ($U/2D=0.2$) we obtain the standard behavior of
temperature dependence of paramagnetic critical field of BCS theory
\cite{S-Gam_Sarma}. At low temperatures we observe the region of decreasing
$H_{cp}$ as temperature diminishes, with maximum $H_{cp}$ at finite temperature.
It is well known, that in this region the system is unstable with respect to
type I phase transition \cite{S-Gam_Sarma}, where is also a possibility of
transition to FFLO phase \cite{FFLO1,FFLO2} with Cooper pairs with finite
momentum (${\bf q}\neq 0$) and inhomogeneous superconducting order parameter.
Critical field in BCS limit is relatively weakly dependent on the value of
chemical potential, so that we can neglect weak field and temperature dependence
of $\mu(H,T)$ (dashed line in Fig. \ref{fig1} (a) in fact coincides with the
result of an exact calculation). With the growth of the coupling strength the
region of instability towards type I transition shrinks (cf. Fig.\ref{fig1}
(b),(c)) and it completely disappears with further increase of coupling
(Fig. \ref{fig1} (d),(e)). With the increase of coupling strength the critical
magnetic field becomes more and more dependent on the value of the chemical
potential, so that the account of its temperature and magnetic field dependence
$\mu(H,T)$ becomes very important (cf. Fig. \ref{fig1} (c-e)).

At intermediate coupling ($U/2D=0.6$) the account of temperature and magnetic
field dependence of $\mu$ leads to small changes of the critical field, however
we observe significant qualitative changes for $T\sim T_c$. The small growth of
chemical potential with increase of $H$ at weak fields leads to noticeable
growth of $T_c$, which overcomes the decrease of $T_c$ with the growth of
magnetic field due to explicit $H$ -- dependence in Eq. (\ref{Hcp}), leading to
some increase of $T_c(H)$  at small $H$.

In Fig. \ref{fig1}(f) сwe show temperature dependencies of the critical magnetic
field for different values of $U$. It is known that the critical temperature
$T_{c0}$ grows with coupling strength in BCS limit and decreases in BEC strong
coupling limit, passing through a maximum at $U/2D=1$ \cite{JETP14,JTL14,JETP15}.
The critical magnetic field at low temperatures grows with coupling strength
both in BCS  and  BEC limits, though in BCS -- BEC crossover region
($U/2D=(0.6-1)$) we observe rather weak dependence of the critical magnetic field
on coupling strength.

\begin{figure}
\includegraphics[clip=true,width=0.3\textwidth]{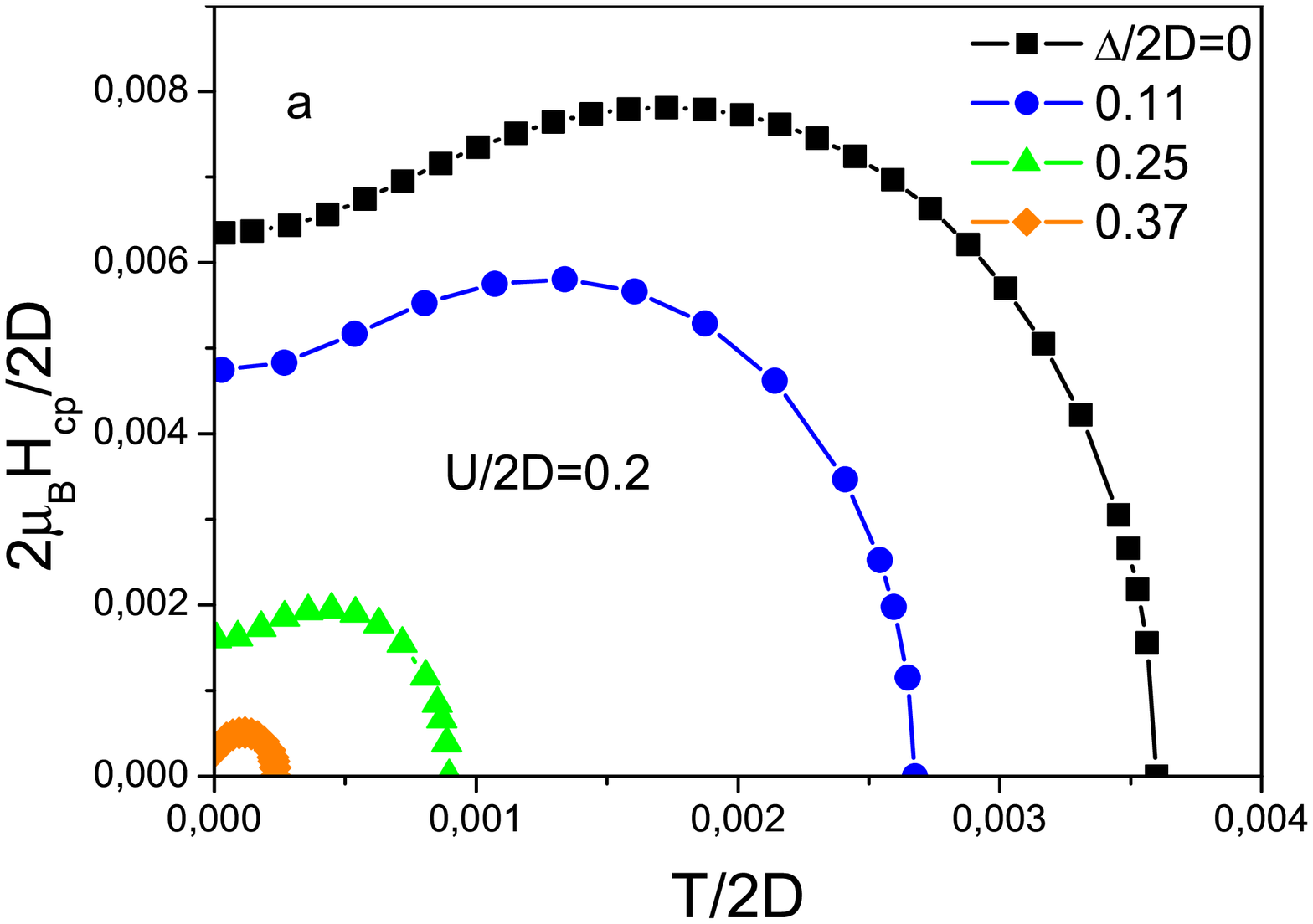}
\includegraphics[clip=true,width=0.27\textwidth]{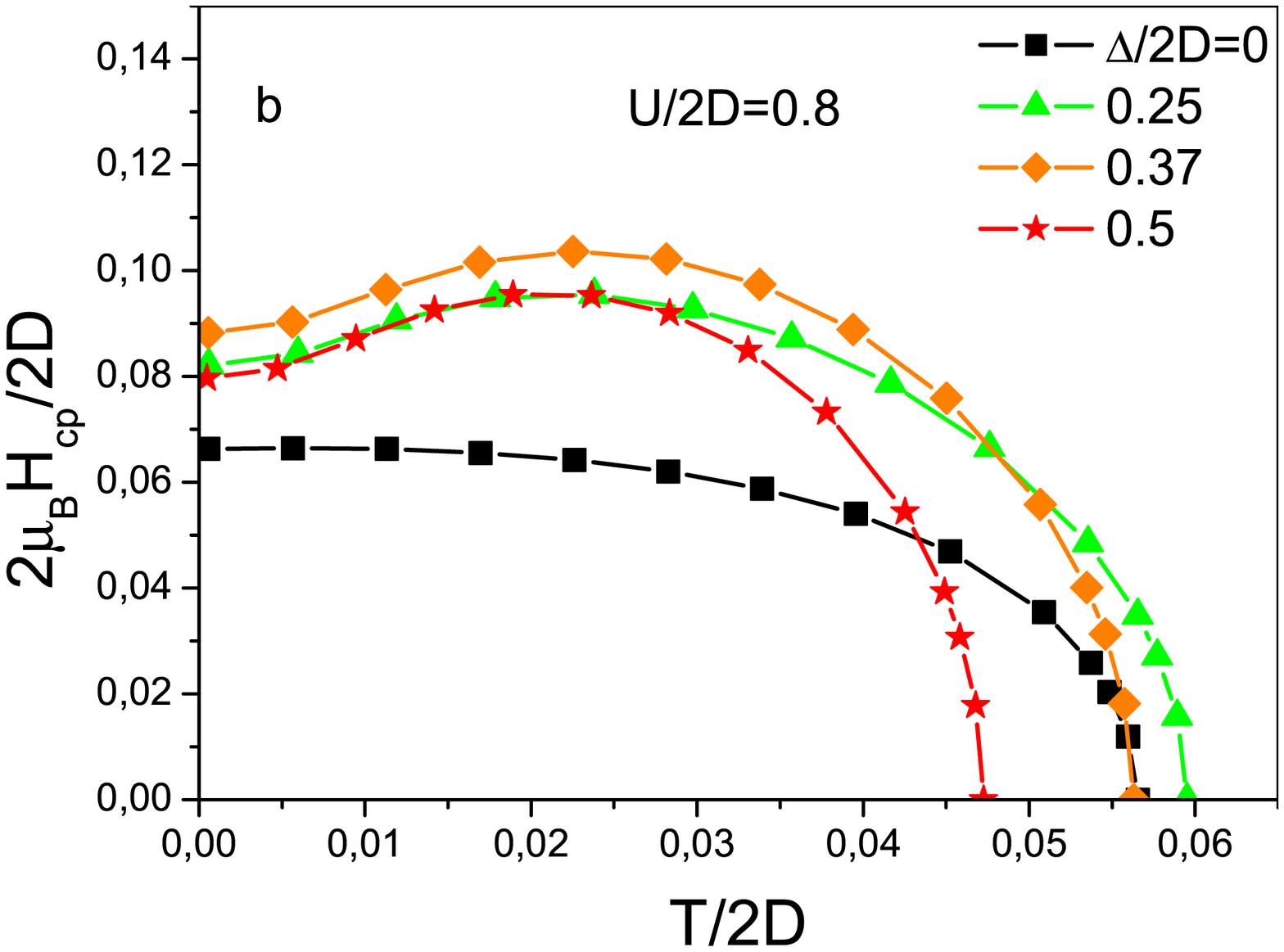}
\includegraphics[clip=true,width=0.27\textwidth]{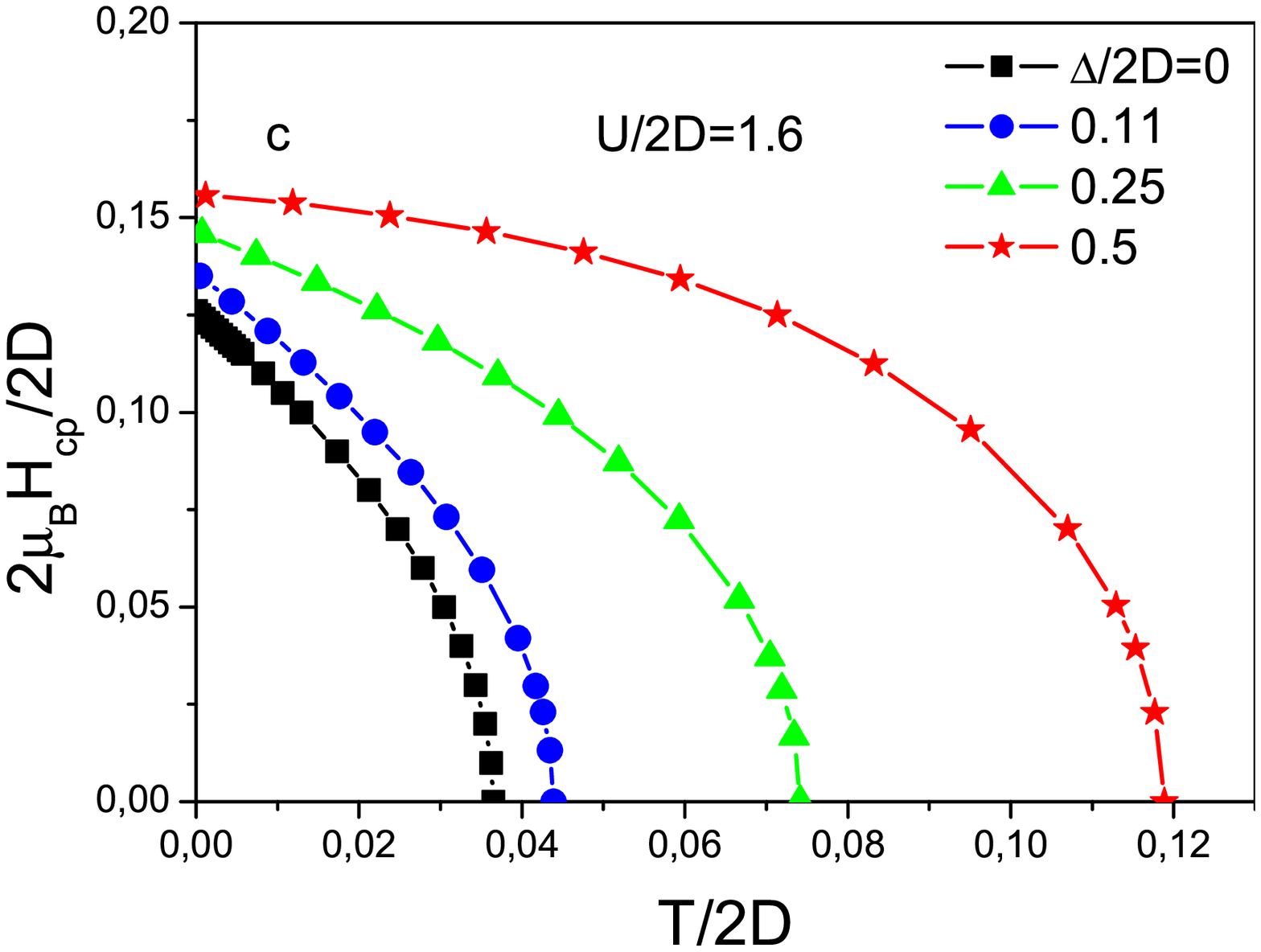}
\caption{Temperature dependence of paramagnetic critical magnetic field for
different levels of disorder:
(a) --- BCS weak coupling limit ($U/2D=0.2$);
(b) --- BCS -- BEC crossover region (intermediate coupling: $U/2D=0.8$);
(c) --- BEC strong coupling region ($U/2D=1.6$).}
\label{fig2}
\end{figure}

The physical reason of the growth of paramagnetic critical field with coupling
strength is pretty obvious --- it is more difficult for magnetic field to break
the pairs of strongly coupled electrons.

In Fig. \ref{fig2} we present our results on disorder influence on temperature
dependence of paramagnetic critical magnetic field. In BCS weak coupling limit
(Fig. \ref{fig2}(а)) the increase of disorder leads both to decrease of the
critical temperature in the absence of magnetic field $T_{c0}$
(cf. \cite{JTL14,JETP15}) and to decrease of the critical magnetic field at all
temperatures. The region of instability to type I transition is conserved also
in the presence of disorder.  In fact, as was noted above, disorder influence
on $H_{cp}(T)$ is actually universal and related only to the replacement
$D\to D_{eff}$. As a result, disorder growth leads to decrease of the effective
coupling, which is defined by dimensionless parameter $U/2D_{eff}$.
This leads to the increase of the relative width $T/T_{c}(H)$ of the temperature
region of type I transition.

At intermediate coupling ($U/2D=0.8$) in BCS -- BEC transition region
(Fig. \ref{fig2}(b)) disorder growth relatively weakly changes the critical
temperature $T_{c0}$ (cf. \cite{JTL14,JETP15}), leading to some increase of
$H_{cp}(T)$. As all the effects of disordering are due to the replacement
$D\to D_{eff}$, the increase of disorder again leads to the decrease of the
effective coupling strength $U/2D_{eff}$ and restoration of the region of
instability towards type I transition.

\begin{figure}
\includegraphics[clip=true,width=0.35\textwidth]{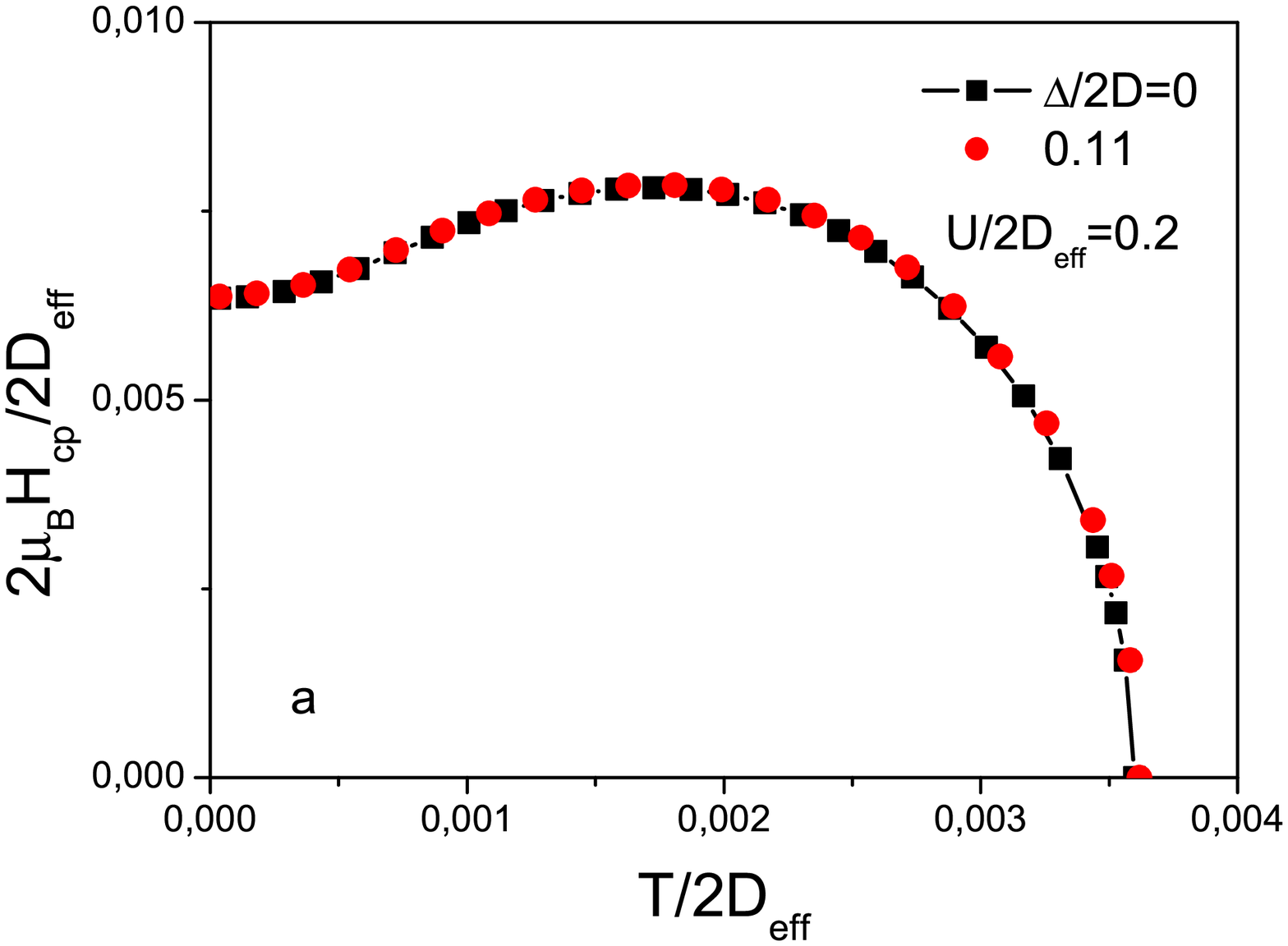}
\includegraphics[clip=true,width=0.35\textwidth]{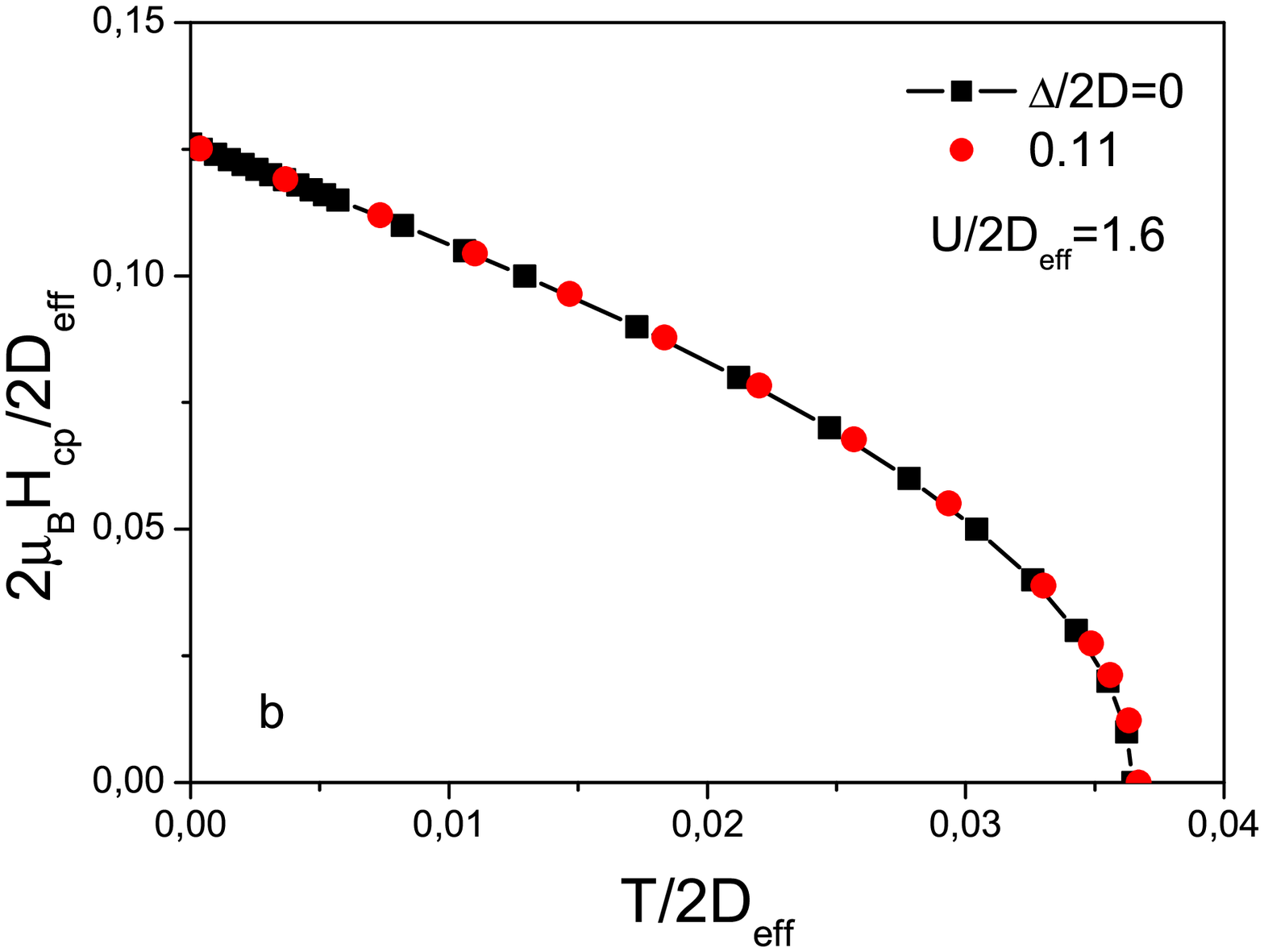}
\caption{Universality of temperature dependence of paramagnetic critical magnetic
field on disorder.
(a) --- weak coupling $U/2D_{eff}=0.2$, $\Delta=0$ и $\Delta=0.11$
(b) --- strong coupling $U/2D_{eff}=1.6$, $\Delta=0$ и $\Delta=0.11$}
\label{fig3}
\end{figure}

In BEC -- limit of strong coupling the growth of disorder leads to significant
increase of the critical temperature $T_{c0}$ (cf. \cite{JTL14,JETP15}).
At the same time, the critical magnetic field at low temperatures only weakly
increases with increasing disorder. In BEC -- limit instability to type I
transition does not appear even in the presence of very strong disorder
($\Delta/2D=0.5$). In fact, in BEC -- limit disorder influence is again
universal and related only to the replacement $D\to D_{eff}$.
As a result, if we make the spin splitting and temperature dimensionless
dividing both by the effective bandwidth $2D_{eff}$ and keep the effective
coupling strength $U/2D_{eff}$ fixed, we obtain the universal temperature
dependence of paramagnetic critical magnetic field. In Fig. \ref{fig3} we show
examples of such universal behavior for typical cases of weak and strong
coupling an the absence and in the presence of disorder.

In the absence of disorder in BEC strong coupling limit with $U/2D=1.6$ for
$T\to 0$ we have (cf. Fig.\ref{fig1}) $2\mu_{B}H_{cp}/2D\approx 0.125$, so that
for characteristic value of the bandwidth $2D\sim 1$ eV we get $H_{cp}\sim
10^{7}$ Gauss. For orbital critical magnetic field (cf. \cite{JETP17_Hc2})
in the same model and for the same coupling strength, for $T\to 0$ and typical
value of lattice parameter $a=3.3*10^{-8}$ cm, we obtain
$H_{cp}\approx 1.6*10^{8}$ Gauss. Thus, the orbital critical magnetic field
at low temperatures grows with increase of the coupling strength much faster,
than paramagnetic critical field, and in BEC strong coupling limit the main
contribution to the upper critical field at low temperatures is actually
due to the paramagnetic effect. The growth of disorder leads to significant
growth of the orbital critical magnetic field \cite{JETP17_Hc2}, while
$H_{cp}(T\to 0)$ in the region of BCS -- BEC crossover and in BEC limit is
relatively weakly dependent on disorder. Thus, also in the presence of disorder
in BEC limit the main contribution to the upper critical field at low temperatures
comes from paramagnetic effect.

\section{Conclusion}

In this paper, within the combination of Nozieres -- Schmitt-Rink and
DMFT+$\Sigma$ approximations, we have studied disorder influence on temperature
behavior of paramagnetic critical magnetic field. Calculations were done for a
wide range of the values of attractive potential $U$, from the weak coupling
region $U/2D\ll 1$, where superconductivity is well described by BCS model,
up to the limit of strong coupling $U/2D\gg 1$, where superconducting transition
is due to Bose condensation of compact Cooper pairs, which are formed at
temperatures much exceeding the temperature of superconducting transition.

The growth of coupling strength $U$ leads to a fast increase of $H_{cp}(T)$
and disappearance, both in the region of BCS -- BEC crossover and in BEC limit,
of the region of instability, leading to type I transition, which appears at low
temperatures in BCS weak coupling region. Physically this is due to the fact,
that it becomes more and more difficult for magnetic field to break pairs of
strongly coupled electrons.

The growth of disorder in BCS weal coupling limit leads both to decrease of
critical temperature and decrease of $H_{cp}(T)$. The region of instability to
type I transition at low temperatures remains also in the presence of disorder.
In the intermediate coupling region ($U/2D=0.8$) disorder only weakly affects
both the critical temperature and $H_{cp}(T)$. However, the growth of disorder
leads to restoration of the low temperature region of instability to type I
transition, which is not observed in the absence of disorder. This, rather
unexpected, conclusion is related to specifics of the attractive Hubbard model,
which in disordered case is controlled by dimensionless coupling parameter
$U/2D_{eff}$. As was shown in our previous works, in BEC strong coupling limit
the growth of disorder leads to noticeable growth of the critical temperature
in the absence of magnetic field. However, the value of $H_{cp}(T\to 0)$ in this
model is relatively weakly dependent on disorder. In BEC limit at low temperatures
and for reasonable values of model parameters paramagnetic critical magnetic
field is much smaller, than the orbital critical field, so that the upper
critical field in this region is mainly determined by paramagnetic critical
filed. In the presence of disorder this conclusion is even more valid, as
the orbital critical field rapidly grows with increasing disorder, while
paramagnetic critical field is weakly disorder dependent in this limit.

This work was performed under the State Contract No. 0389-2014-0001 with
partial support of RFBR Grant No. 17-02-00015 and the Program of Fundamental
Research of the RAS Presidium No. 12
 ``Fundamental problems of high -- temperature superconductivity''.

\appendix
\section{Appendix: Equation for paramagnetic critical magnetic field}

In general case the Noziers -- Schmitt-Rink approach \cite{NS} assumes, that
corrections from strong pairing interaction significantly change the chemical
potential of the system, but possible vertex corrections from this interaction
in Cooper channel are irrelevant, so that to analyze Cooper instability we can
use the weak coupling approximation (ladder approximation). In this approximation
the condition of Cooper instability in disordered attractive Hubbard model is
written as:
\begin{equation}
1=U\chi_0(q=0,\omega_m=0)
\label{Cupper}
\end{equation}
where
\begin{equation}
\chi_0(q=0,\omega_m=0)=T\sum_{n}\sum_{\bf pp'}\Phi_{\bf pp'}(\varepsilon_n)
\label{chi}
\end{equation}
is two -- particle loop in Cooper channel ``dressed'' only by impurity
scattering, while $\Phi_{\bf pp'}(\varepsilon_n)$ is the averaged over impurities
two -- particle Green's function in Cooper channel at Matsubara frequencies
$\varepsilon_n=\pi T(2n+1)$.

To obtain $\sum_{\bf pp'}\Phi_{\bf pp'}(\varepsilon_n)$ we use an exact Ward
identity, derived by us in Ref. \cite{PRB07}:
\begin{eqnarray}
G_\uparrow (\varepsilon_n,{\bf p})-G_\downarrow (-\varepsilon_n,-{\bf p})=
\nonumber\\
-\sum_{\bf p'}\Phi_{\bf pp'}(\varepsilon_n)
(G_{0\uparrow}^{-1}(\varepsilon_n,{\bf p'})-G_{0\downarrow}^{-1}
(-\varepsilon_n,-{\bf p})),
\label{Word}
\end{eqnarray}
Here $G_{0\uparrow,\downarrow}(\varepsilon_n,{\bf p})=
(i\varepsilon_n+\mu-\varepsilon({\bf p})\pm \mu_BH)^{-1}$
is the ``bare'' Green's function and
$G_{\uparrow ,\downarrow}(\varepsilon_n,{\bf p})$ is averaged over impurities
(but not ``dressed'' by Hubbard interaction!) single -- particle Green's
function. Using the symmetry $\varepsilon({\bf p})=\varepsilon(-{\bf p})$ we
obtain from Ward identity (\ref{Word}):
\begin{equation}
\sum_{\bf pp'}\Phi_{\bf pp'}(\varepsilon_n)=
-\frac{\sum_{\bf p}G_\uparrow (\varepsilon_n,{\bf p})-
\sum_{\bf p}G_\downarrow (-\varepsilon_n,{\bf p})}{2i\varepsilon_n+2\mu_BH},
\label{Phi}
\end{equation}
so that for Cooper susceptibility (\ref{chi}) we get:
\begin{eqnarray}
\chi_0(q=0,\omega_m=0)=\nonumber\\
-\frac{T}{2}\sum_{n}\frac{\sum_{\bf p}G_\uparrow (\varepsilon_n,{\bf p})-
\sum_{\bf p}G_\downarrow (-\varepsilon_n,{\bf p})}{i\varepsilon_n+\mu_BH}=\nonumber\\
=-\frac{T}{2}\sum_{n}\left( 
\frac{\sum_{\bf p}G_\uparrow(\varepsilon_n,{\bf p})}{i\varepsilon_n+\mu_BH}+
\frac{\sum_{\bf p}G_\downarrow(\varepsilon_n,{\bf p})}{i\varepsilon_n-\mu_BH}
\right) .
\label{chi1}
\end{eqnarray}
Performing the standard summation over Fermion Matsubara frequencies, we obtain:
\begin{eqnarray}
\chi_0=-\frac{1}{8\pi i}\int_{-\infty}^{\infty}d\varepsilon
\left(
\frac{\sum_{\bf p}G_\uparrow^R(\varepsilon,{\bf p})-
\sum_{\bf p}G_\uparrow^A(\varepsilon,{\bf p})}{\varepsilon+\mu_BH}+\right.\nonumber\\
+\left.\frac{\sum_{\bf p}G_\downarrow^R(\varepsilon,{\bf p})-
\sum_{\bf p}G_\downarrow^A(\varepsilon,{\bf p})}{\varepsilon-\mu_BH}
\right)th\frac{\varepsilon}{2T}=\nonumber\\
=\frac{1}{4}\int_{-\infty}^{\infty}d\varepsilon
\left(
\frac{\tilde N_{0\uparrow}(\varepsilon)}{\varepsilon +\mu_BH}+
\frac{\tilde N_{0\downarrow}(\varepsilon)}{\varepsilon -\mu_BH}
\right)
th\frac{\varepsilon}{2T},\nonumber\\
\label{chi2}
\end{eqnarray}
where $\tilde N_{0\uparrow ,\downarrow}(\varepsilon)$ is the ``bare'' ($U=0$)
density of states for different spin projections, ``dressed'' by impurity
scattering. Spin splitting can be considered as an addition to chemical
potential, so that introducing the ``bare'' density of states ``dressed'' by
disorder in the absence of external magnetic field $\tilde N_0(\varepsilon)$,
we obtain the final result for Cooper susceptibility:
\begin{equation}
\chi_0=\frac{1}{4}\int_{-\infty}^{\infty}d\varepsilon
\frac{\tilde N_0(\varepsilon)}{\varepsilon}
\left(
th\frac{\varepsilon +\mu_BH}{2T}+th\frac{\varepsilon -\mu_BH}{2T}
\right)
\label{chi3}
\end{equation}
In Eq. (\ref{chi3}) energy $\varepsilon$ is counted from the chemical potential
level. If we count it from the middle of the conduction band we have to
replace $\varepsilon\to\varepsilon -\mu$ and the condition of Cooper instability
(\ref{Cupper}) leads to the equation defining critical temperature depending on
the external magnetic field, which gives the equation for paramagnetic critical
magnetic filed (\ref{Hcp}).  The chemical potential for different values of
$U$ and $\Delta$ should be determined from DMFT+$\Sigma$ calculations, i.e. from
the standard equation for electron number (band filling), which allows us to
find $H_{cp}$ for the wide range of model parameters, including the region of
BCS -- BEC crossover and the limit of strong coupling at different levels of
disorder. This reflects the physical meaning of Nozieres -- Scmitt-Rink
approximation --- in the weak coupling region the temperature of superconducting
transition is controlled by the equation for Cooper instability (\ref{Hcp}),
while in the strong coupling limit it is defined as the temperature of BEC, which
is controlled by chemical potential. The joint solution of Eq. (\ref{Hcp}) and
the equation for the chemical potential guarantees the correct interpolation for
$H_{cp}$ in the region of BCS -- BEC crossover.


\newpage


\begin{thebibliography}{99}

\bibitem{NS} P. Nozieres and S. Schmitt-Rink, J. Low Temp. Phys. {\bf 59}, 195
(1985).
\bibitem{pruschke}  Th. Pruschke, M. Jarrell, J. K. Freericks. Adv. Phys.
{\bf 44}, 187 (1995).
\bibitem{georges96}  A. Georges, G. Kotliar, W. Krauth, M. J. Rozenberg.
Rev. Mod. Phys. {\bf 68}, 13 (1996).
\bibitem{Vollh10}D. Vollhardt in ``Lectures on the Physics of Strongly
Correlated Systems XIV'', eds. A. Avella and F. Mancini, AIP Conference
Proceedings vol. 1297 (AIP, Melville, New York, 2010), p. 339; ArXiV: 1004.5069.
\bibitem{JTL05}E.Z.Kuchinskii, I.A.Nekrasov, M.V.Sadovskii.
Pisma Zh. Eksp. Teor. Fiz. {\bf 82}, 217 (2005) [JETP Letters {\bf 82}, 198 (2005)];
ArXiv: cond-mat/0506215.
\bibitem{PRB05}M.V. Sadovskii, I.A. Nekrasov, E.Z. Kuchinskii, Th. Prushke,
V.I. Anisimov. Phys. Rev. B {\bf 72}, No 15, 155105 (2005);
ArXiV: cond-mat/0508585.
\bibitem{FNT06}E.Z. Kuchinskii, I.A. Nekrasov, M.V. Sadovskii. 
Fizika Nizkih Temperatur {\bf 32}, 528 (2006) [Low Temp. Phys. {\bf 32}, 398 (2006)];
ArXiv: cond-mat/0510376.
\bibitem{PRB07}E.Z. Kuchinskii, I.A. Nekrasov, M.V. Sadovskii.
Phys. Rev. B {\bf 75}, 115102-115112 (2007); ArXiv: cond-mat/0609404.
\bibitem{UFN12}E.Z. Kuchinskii, I.A. Nekrasov, M.V. Sadovskii.
Usp. Fiz. Nauk {\bf 182}, 345 (2012) [Physics Uspekhi {\bf 53}, 325 (2012)];
ArXiv:1109.2305.
\bibitem{HubDis} E.Z. Kuchinskii, I.A. Nekrasov, M.V. Sadovskii,
Zh. Eksp. Teor. Fiz. {\bf 133}, 670 (2008) [JETP {\bf 106}, 581 (2008)];
ArXiv: 0706.2618.
\bibitem{LVK16}E.Z. Kuchinskii, M.V. Sadovskii. Zh. Eksp. Teor. Fiz.  {\bf 149},
589 (2016) [JETP {\bf 122}, 509 (2016)]; ArXiv:1507.07654
\bibitem{JETP14}N.A. Kuleeva, E.Z. Kuchinskii, M.V. Sadovskii.
Zh. Eksp. Teor. Fiz. {\bf 146}, 304  (2014) [JETP {\bf 119}, 264 (2014)];
ArXiv: 1401.2295.
\bibitem{JTL14} E.Z. Kuchinskii, N.A. Kuleeva, M.V. Sadovskii.
Pisma Zh. Eksp. Teor. Fiz. {\bf 100}, 213  (2014)
[JETP Letters {\bf 100}, 192  (2014)]; ArXiv: 1406.5603.
\bibitem {JETP15} E.Z. Kuchinskii, N.A. Kuleeva, M.V. Sadovskii.
Zh. Eksp. Teor. Fiz. {\bf 147}, 1220  (2015)
[JETP {\bf 120}, 1055 (2015)]; ArXiv:1411.1547.
\bibitem{JETP17_Hc2} E.Z. Kuchinskii, N.A. Kuleeva, M.V. Sadovskii.
Zh. Eksp. Teor. Fiz. {\bf 152}, 1321 (2017)
[ JETP 125, No.6, 1127 (2017)]; ArXiv:1709.03895
\bibitem{FFLO1} P. Fulde, R.A. Ferrell. Phys. Rev., {\bf A135}, 550 (1964)
\bibitem{FFLO2} A.I. Larkin, Yu.N. Ovchinnikov. Zh. Eksp. Teor. Fiz. {\bf 47},
1136 (1964)
\bibitem{S-Gam_Sarma} D. Saint-James, G. Sarma, E.J. Thomas.
Type II Superconductivity. Pergamon Press, Oxford, 1969
\bibitem{NRGrev} R. Bulla, T.A. Costi, T. Pruschke, Rev. Mod. Phys. {\bf 60},
395 (2008).


\end{thebibliography}
\end{document}